# Efficient generation of large-scale non-equilibrium distributions of particles

Sergejs Tarasovs

Institute for Mechanics of Materials, Faculty of Science and Technology, University of Latvia, Jelgavas St. 3, LV-1004, Riga

E-mail: sergejs.tarasovs@lu.lv

**Abstract**

This work presents an efficient algorithm for generating statistically representative microstructures of particulate composites in periodic representative volume elements. The Swelling and Random Migration (SRM) algorithm combines collective particle rearrangements with an adaptive cell-based neighbor search scheme, enabling near-linear computational scaling for low to intermediate volume fractions and allowing simulations with up to $10^7$ particles in two and three dimensions. SRM offers great flexibility, allowing the controlled generation of both equilibrium-like and strongly non-equilibrium particle arrangements. The method is readily extendable to non-spherical inclusions; this capability is demonstrated by modeling thin circular platelets and generating qualitatively distinct platelet microstructures, including highly interconnected "house-of-cards" networks and metastable quasi-nematic domains. The results highlight the importance of microstructural arrangement in structure-property relationships and establish SRM as a powerful tool for generating realistic, diverse, and computationally accessible particle configurations for composite material modeling.

## 1. Introduction

Representative volume elements (RVEs) with periodic microstructures play an important role in physics, materials science, and engineering. They provide a framework for linking microscale heterogeneity with macroscale effective properties, enabling systematic investigation of transport, mechanical, thermal, and electromagnetic behavior in heterogeneous media. In physics, the generation of such particle configurations is traditionally guided by equilibrium statistical mechanics or by extremal packing states, such as random close packing, maximally jammed configurations, or energy-minimized ensembles. As a result, many established algorithms, such as Monte Carlo algorithms [1], Lubachevsky–Stillinger-type event-driven molecular dynamics [2, 3], and collective rearrangement methods [4-6], are optimized to reproduce equilibrium or near-equilibrium structures.

In engineering and material science, manufacturing processes such as filament winding, pultrusion, injection molding, or powder compaction produce particle distributions that are non-uniform, anisotropic, and often history-dependent. Unidirectional fiber-reinforced composites exhibit spatial correlations induced by flow and curing; particulate composites display clustering, segregation, or alignment driven by processing conditions; and hybrid systems may combine particles of different shapes, sizes, and interactions. These features cannot be captured reliably by algorithms designed to approximate equilibrium states, as such methods tend to suppress the very irregularities, defects, and controlled disorder that characterize real composite microstructures.

Consequently, there is a growing need for new algorithms capable of producing periodic RVEs with tunable degrees of randomness, controlled spatial correlations, and realistic non-equilibrium features. Such methods should allow researchers to systematically vary microstructural disorder, mimic manufacturing-induced anisotropy, and generate ensembles of RVEs that reflect the statistical variability observed in practice. Achieving this requires computationally efficient frameworks that can reproduce a broad spectrum of particle arrangements, from highly ordered to strongly disordered, while maintaining periodicity and avoiding numerical artifacts.

It is important to emphasize that the notion of *non-equilibrium* used in this study differs from its thermodynamic meaning. Here, non-equilibrium configurations refer to microstructures exhibiting clustering, agglomeration, or domain-like organization of particles, features frequently encountered in particulate composites and colloidal materials due to processing conditions, surface interactions, or fabrication constraints. These structures are *geometrically* non-uniform but are not associated with thermodynamic driving forces or kinetic pathways.

Different approaches have been used to generate RVEs with random distributions of non-penetrating spheres or disks in 3D or 2D space, respectively. Metropolis algorithm [1] has often been used to generate equilibrium configurations of hard spheres or disks, where particles are initially placed in a periodic array, and each particle is moved by a small distance to a new position, controlling overlapping with neighboring particles. The process is continued iteratively until the equilibrium configuration is reached. Generation of equilibrium configurations becomes difficult for systems near the freezing limit; therefore, the "swelling" of particles was used in [7] for high volume fractions. In this approach, an initial equilibrium configuration was generated by Metropolis' algorithm, and then all particles are allowed to "swell" (i.e., increase the volume), and if overlapping is observed, overlapping particles are moved by a small distance until a non-overlapping configuration is found. If, after several attempts, no such configuration is found, the entire system is "shaken" by randomly moving each particle.

An alternative strategy, the Modified Random Sequential Adsorption (RSA) algorithm, was proposed by Segurado and Llorca [8], where the RSA algorithm was used to generate an initial configuration, and RVE shrinking with random perturbation of overlapping particles was used to achieve higher volume fractions of inclusions. Dense packing (jamming limit) of particles can be generated by an adaptive shrinking cell algorithm [9].

Lubachevsky–Stillinger [2] approach is an event-driven or collision-driven molecular-dynamics algorithm where an initial sparse configuration of particles is given initial random velocities, and the motion of particles is followed as they collide and grow in volume until the particles can no longer expand. Depending on the particles' growth rate, a system with high density can be generated; at lower densities, the algorithm produces an equilibrium distribution of particles.

Special algorithms were developed in composite mechanics studies, where generated configurations should mimic non-uniform distributions observed in practice. Modifications to the molecular dynamics algorithm were proposed [10] that allow for the generation of configurations of fibers with different amounts of clusters and resin-rich zones, statistically equivalent to real microstructures. A fiber rearrangement procedure was used in [11] to generate synthetic microstructures that resemble real UD composites by comparison of nearest neighbor and radial distribution statistical descriptors. An experimentally measured 1st and 2nd Nearest Neighbor distribution functions were used in [12] to generate statistically equivalent fiber distributions. A particle swarm optimization was used in [13] for the generation of microstructures with different volume fractions and statistical descriptors similar to real composites. Specific algorithms were developed for generating stochastic distributions with distinct cluster of fibers [14, 15].

In the Random Sequential Expansion (RSE) algorithm [16], new particles are sequentially added to the RVE around the first particle placed in the center, resulting in a compact cluster of fibers around the initial fiber, and then iteratively adding new fibers around this cluster. However, the volume fraction of fibers and the distribution of fibers in the RVE are difficult to control. To overcome this problem, the Random Fiber Removal algorithm was proposed in [17], where an initial dense configuration was generated using the RSE algorithm, and configurations with lower densities were obtained by removing random fibers from the initial configuration. Melro et al. [18] have proposed a modification to the RSA algorithm, where fibers are shifted at each iteration using several complex heuristic algorithms to free space for new fibers.

In another class of algorithms, an initially overlapping configuration of particles with a desired volume fraction is used with a subsequent iterative optimization process of particles shifting subject to a certain cost function [19-21] until a non-overlapping configuration is found.

The development of such algorithms is essential for advancing predictive modeling of composite materials. By enabling the generation of realistic microstructures with controllable randomness, these methods support more accurate homogenization studies, uncertainty quantification, virtual testing, and data-driven materials design. The present work proposes a Swelling and Random Migration (SRM) algorithm for efficient generating of large-scale periodic random particle distributions tailored to the requirements of composite material modeling.

## 2. Swelling and Random Migration algorithm

In this work, statistically representative volume elements containing randomly distributed monodisperse inclusions are generated using a hybrid approach that combines the modified Miller–Torquato algorithm [7] with a modified Random Sequential Addition (RSA) scheme [8]. The procedure begins by placing the required number of inclusions at a low initial volume fraction inside a unit box using the classical RSA algorithm [22, 23]. Subsequently, all particles are iteratively expanded until the target volume fraction is reached. Each iteration consists of the following steps:

a) Swelling: all particles are increased in radius by a small increment.

b) Random migration: each particle is displaced by a small random vector.

c) Local admissibility check: for every particle, distances to its neighbors are evaluated. If the minimum distance falls below a prescribed threshold, the particle is returned to its previous position, and a new random displacement is attempted. This process continues until a non-overlapping configuration is found.

d) Shake step: if no admissible position can be found for any particle after several attempts, all changes from the current iteration are discarded, and a global relaxation step is performed:

i. each particle is randomly displaced and tested for overlap; if a valid position is found, it is accepted, otherwise the particle is restored, and a second attempt is made;

ii. after a fixed number of unsuccessful attempts (or after a successful relocation), the procedure advances to the next particle;

iii. several cycles of this shake procedure are executed, and the resulting configuration is accepted as the starting point for the next iteration.

e) Termination: the iterative swelling–migration–shake cycle continues until the prescribed inclusion volume fraction is achieved.

Throughout the process, periodicity is strictly enforced either by the minimum-image convention [24, 25], or by direct generation of image particles: after each particle translation, boundary crossings are detected, and the corresponding periodic images are added or removed to maintain geometric consistency [26].

The global shaking step inevitably disrupts compact, dense clusters; however, it plays a crucial role in overcoming local jamming, which frequently arises at low particle volume fractions during the generation of intentionally clustered microstructures. With an appropriate choice of swelling and migration rates, the algorithm efficiently produces configurations ranging from equilibrium-like to strongly clustered arrangements, achieving volume fractions up to the random close-packing limit for disks and approximately 60% for spherical particles. For higher volume fractions, alternative approaches, such as event-driven molecular dynamics, remain more effective.

For large RVEs, the most computationally demanding component of the algorithm is the detection of overlapping particles. In a naïve implementation, the number of pairwise distance checks grows quadratically with the number of inclusions, making the procedure prohibitively expensive for large-scale microstructures. This bottleneck can be eliminated by introducing a cell-based spatial partitioning scheme, following the classical approach of Clarke and Willey [4].

In this method, the RVE domain is subdivided into small cells, each storing a list of particles whose centers lie within that cell (Fig. 1). By selecting the cell size slightly larger than the particle diameter, each cell contains only a few inclusions. Consequently, for any given particle, potential overlaps need to be checked only against particles located in the nine neighboring cells (highlighted in Fig. 1). The average number of candidate neighbors (indicated by red and orange colors in Fig. 1) remains independent of the total number of particles in the RVE. As a result, the overall complexity of collision detection becomes linear with respect to the system size, enabling efficient generation of large RVEs with tens or hundreds of thousands of inclusions.

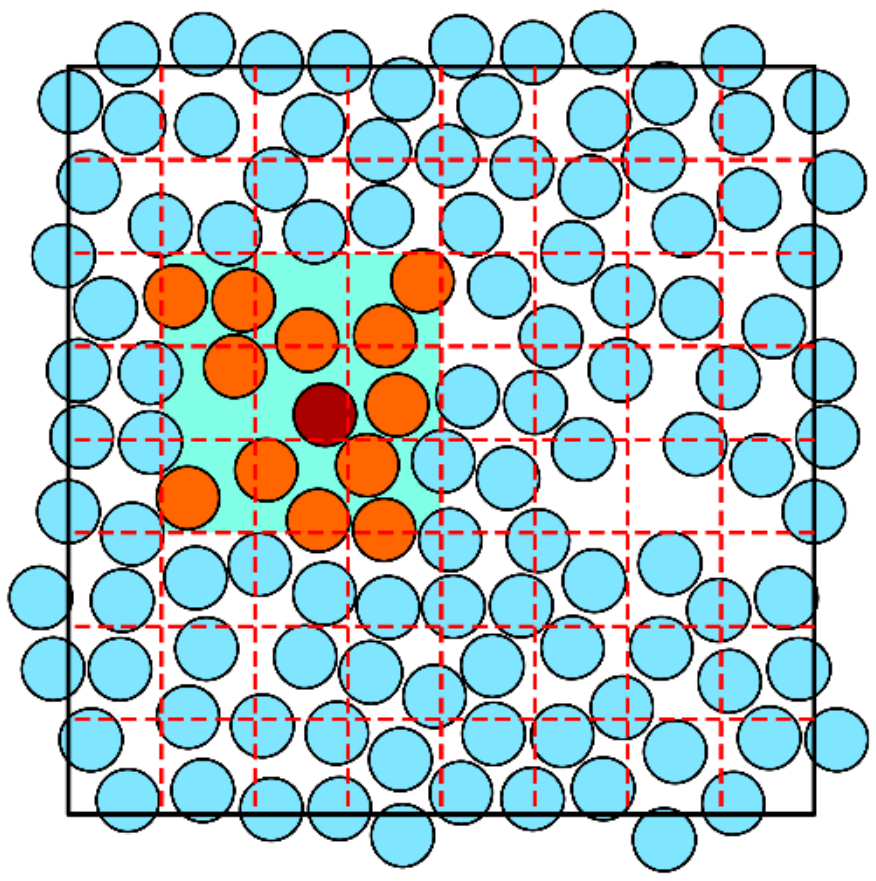

Fig. 1. Cell-based search procedure for detecting overlapping particles.

A pseudocode representation of the basic two-dimensional SRM algorithm is provided below; its extension to three dimensions is straightforward. This formulation incorporates a cell-based neighbor search for efficient collision detection, while periodic images of particles are handled using the minimum-image convention. The procedure begins with the generation of a non-overlapping initial configuration of particles within a rectangular domain using the classical RSA algorithm; this initialization step is not detailed here. The initial particle volume fraction is denoted by $f_0$, and the algorithm terminates once the volume fraction $f$ exceeds the prescribed target value $f_{target}$. Minor adjustments may be required during the final iteration to match the target volume fraction exactly.

Two user-defined parameters, the swelling rate $c_w$ and the migration rate $c_m$, govern the degree of randomness in the resulting particle configuration. Two additional parameters, $N_k$ and $N_l$, specify the number of attempts allowed when searching for a new non-overlapping position for each particle; values on the order of 5–10 are typically sufficient. The SRM algorithm employs two nested search loops when resolving particle overlaps. Although not strictly necessary, this structure significantly enhances robustness and helps the algorithm escape local jamming, which is common in clustered configurations and at high volume fractions.

Algorithm 1 summarizes the main procedure and calls Algorithms 2–6, which implement the individual components of the SRM method.

| Algorithm 1: Main program | |
|---|---|
| 1: | Define a rectangle with sides $L_x$ and $L_y$ |
| 2: | Define a set of non-overlapping particles (disks) $P$ with positions $\mathbf{r}_i$, radiuses $R_i$, and volume fraction $f_0$ (RSA procedure) |
| 3: | **while** $f < f_{target}$ **do** |
| 4: | $cellSize = 2.5 * \max(R_i) + c_m$ |
| 5: | $N_{x,y} = \max\,(3, (\text{int})(L_{x,y}\ /\ cellSize))$ |
| 6: | $cells = Array[N_x, N_y]$ |
| | *// Copy particles* |
| 7: | $Q = P$ |
| 8: | Scale all particles $Q$ by a swelling rate factor $c_w$ |
| 9: | **for** $k$ in $\{1..N_k\}$ **do** |
| 10: | Fill $cells$ using Algorithm 2 |
| 11: | Move particles $Q$ using Algorithm 3 |
| 12: | *collision* = Check collisions for $Q$ using Algorithm 6 |
| 13: | **if** No *collision* **break** |
| 14: | **end for** |
| 15: | **if** No *collision* **then** |
| 16: | $P = Q$ |
| 17: | Update volume fraction of particles $f$ |
| 18: | **else** |
| 19: | Fill $cells$ using Algorithm 2 |
| | *// Shake particles* |
| 20: | **while** $k++ < N_k$ **do** |
| 21: | Move particles $P$ using Algorithm 3 |
| 22: | **end while** |
| 23: | **end if** |
| 24: | **end while** |

Algorithm 2 constructs a two-dimensional cell array in which each cell stores a list of particles whose centers lie within that spatial region. The cell size is updated at every iteration of the main procedure (Algorithm 1) based on the current maximum particle radius and the prescribed migration rate. This adaptive adjustment ensures that each cell contains only a small number of particles, thereby maintaining the efficiency of the neighbor-search procedure throughout the swelling–migration process.

| Algorithm 2: Build cell list | |
|---|---|
| 1: | **for each** particle $q$ in $Q$ **do** |
| 2: | $c_x = (\text{int})(q.X \ / \ cellSize)$ |
| 3: | $c_y = (\text{int})(q.Y \ / \ cellSize)$ |
| 4: | $c_x = (c_x \ \% \ N_x + N_x) \ \% \ N_x$ |
| 5: | $c_y = (c_y \ \% \ N_y + N_y) \ \% \ N_y$ |
| 6: | $cells[c_x, c_y].\text{Add}(q)$ |
| 7: | **end for** |

Algorithm 3 performs a randomized search to identify a new admissible, non-overlapping position for each particle. For every particle, a sequence of small random displacements is generated, and after each trial move the particle is tested for potential overlaps with its neighbors using the cell-based search structure. If the proposed position is collision-free, it is accepted; otherwise, the particle is returned to its previous location, and another random displacement is attempted. This procedure continues until either a valid position is found or the maximum number of allowed attempts is reached.

| Algorithm 3: Move particles | |
|---|---|
| 1: | **for each** particle $q$ in $Q$ **do** |
| 2: |     $q_{old} = q$ |
| 3: |     **for** $l$ in $\{1..N_l\}$ **do** |
| 4: |         $\mathbf{v} = c_m * Unit\ Random\ Vector$ |
| 5: |         $q.X = q_{old}.X + v.X$ |
| 6: |         $q.Y = q_{old}.Y + v.Y$ |
| 7: |         Apply periodic wrapping for $q$ using Algorithm 4 |
| 8: |         *collision* = Check collisions for $q$ using Algorithm 5 |
| 9: |         if No *collision* break |
| 10: |         $q = q_{old}$ |
| 11: |     **end for** |
| 12: | **end for** |

Algorithm 4 enforces periodic boundary conditions by wrapping particles that move outside the simulation domain back into the opposite side of the RVE. This ensures that the particle configuration remains fully consistent with the imposed periodicity throughout the entire swelling–migration procedure.

| Algorithm 4: Periodic wrapping | |
|---|---|
| 1: | if $(q.X < 0)$ $q.X \mathrel{+}= L_x$ |
| 2: | if $(q.X \geq L_x)$ $q.X \mathrel{-}= L_x$ |
| 3: | if $(q.Y < 0)$ $q.Y \mathrel{+}= L_y$ |
| 4: | if $(q.Y \geq L_y)$ $q.Y \mathrel{-}= L_y$ |

Algorithm 5 identifies all potential colliding neighbors of a particle $q$ by searching within the $3 \times 3$ block of cells surrounding the cell that contains $q$. Periodicity of the RVE is enforced using the minimum-image convention, ensuring that neighbors across the domain boundaries are correctly detected and that all overlap checks remain consistent with the imposed periodic boundary conditions.

| Algorithm 5: Collision check for particle $q$ | |
|---|---|
| 1: | $c_x = (\text{int})(q.X\ /\ cellSize)$ |
| 2: | $c_y = (\text{int})(q.Y\ /\ cellSize)$ |
| 3: | $c_x = (c_x\ \%\ N_x + N_x)\ \%\ N_x$ |
| 4: | $c_y = (c_y\ \%\ N_y + N_y)\ \%\ N_y$ |
| 5: | **for** $s_x$ in $\{-1; 0; 1\}$ |
| 6: |     **for** $s_y$ in $\{-1; 0; 1\}$ |
| 7: |         $n_x = (c_x + s_x + N_x)\ \%\ N_x$ |
| 8: |         $n_y = (c_y + s_y + N_y)\ \%\ N_y$ |
| 9: | **for each** particle $t$ in $cells[n_x, n_y]$ |
| 10: | $d_x = q.X - t.X$ |
| 11: | $d_y = q.Y - t.Y$ |
| | // Use "*minimum-image convention*" for periodic particles |
| 13: | **if** $(d_x > 0.5L_x)$ $d_x -= L_x$ |
| 14: | **if** $(d_x < -0.5L_x)$ $d_x += L_x$ |
| 15: | **if** $(d_y > 0.5L_y)$ $d_y -= L_y$ |
| 16: | **if** $(d_y < -0.5L_y)$ $d_y += L_y$ |
| 17: | $S = \sqrt{d_x^2 + d_y^2}$ |
| 18: | **if** $(S \leq q.R + t.R)$ **return** *true* |
| 19: | **end for** |
| 20: | **end for** |
| 21: | **end for** |
| 22: | **return** *false* |

Algorithm 6 performs a collision check for the entire RVE.

| Algorithm 6: Collision check for particles | |
|---|---|
| 1: | **for each** particle $q$ in $Q$ **do** |
| 2: | *collision* = Check collisions for $q$ using Algorithm 5 |
| 3: | **if** *collision* **return** *true* |
| 4: | **end for** |
| 5: | **return** *false* |

The SRM algorithm belongs to the class of collective-rearrangement methods, in which $N$ randomly distributed points are assigned radii and velocities, and all particles are displaced simultaneously as their radii are gradually increased. At each iteration, overlaps are resolved, allowing the algorithm to efficiently generate dense, non-overlapping configurations even at moderate to high volume fractions, provided that an optimized implementation is used. A further advantage of the SRM framework is its simplicity and straightforward extensibility to a wide range of particle shapes. These features make SRM well-suited for generating diverse microstructures, and the following section demonstrates its performance and versatility through several representative examples.

# 3. Results and discussions

## 3.1. Efficiency of SRM algorithm

The efficiency of the SRM algorithm is largely determined by its ability to resolve multiple overlapping particles within a single collective rearrangement step. Although each iteration is computationally more expensive than in event-driven molecular dynamics (MD), the total number of iterations required for convergence remains relatively small and, in practice, is nearly independent of the number of particles. With the incorporation of a cell-based neighbor search, the theoretical computational complexity of the SRM algorithm becomes $O(N)$, since the number of iterations does not grow with system size for low to intermediate volume fractions.

In practical implementations, however, the generation time for the straightforward SRM scheme described above scales approximately as $t \propto N^{1.3}$. This deviation from linear scaling is primarily caused by cache misses arising from the random-access pattern inherent in the cell-list structure. A simple yet effective optimization is to reorder particles access to improve spatial locality and reduce cache-access penalties. This modification yields a 2–3-fold performance improvement and results in a scaling behavior of approximately $t \propto N^{1.15}$ for both two- and three-dimensional systems.

The performance of the optimized algorithm was evaluated on systems containing up to $\sim 10^7$ particles in 2D and 3D. Initial configurations with a volume fraction of 0.1 were generated using the RSA algorithm, after which the SRM algorithm was applied to reach target volume fractions of 0.5 in 2D and 0.4 in 3D. For these benchmarks, relatively large swelling and migration rates were selected to reduce computation times, which are reported in Fig. 2. It should be noted that the actual generation time depends strongly on the chosen SRM parameters; in particular, the creation of clustered microstructures may require significantly longer times than equilibrium-like configurations. The computer codes used in these tests are available at https://github.com/RiSearcher/SRM.

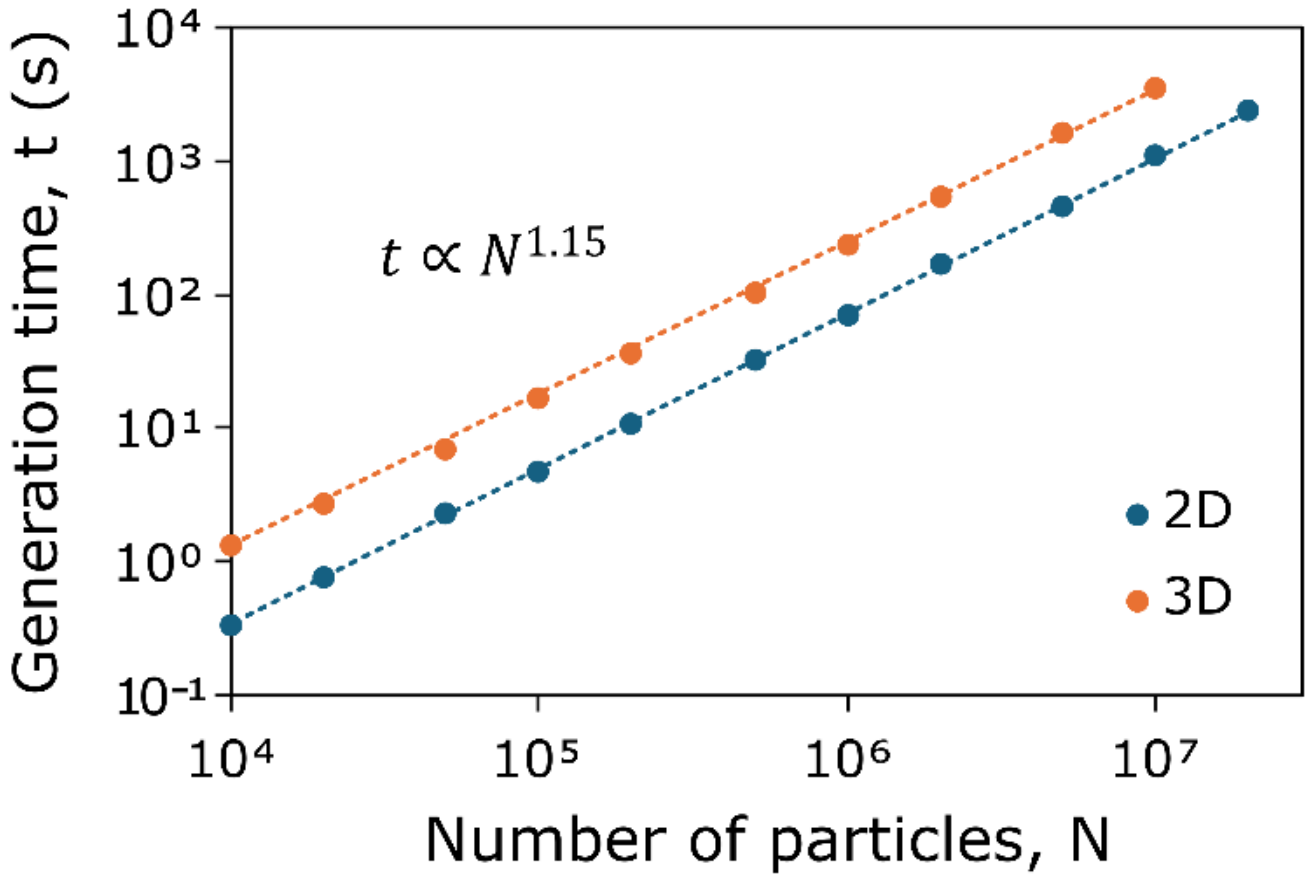

Fig. 2. Generation times of SRM algorithm for RVE with volume fraction of 0.5 for disks and 0.4 for spheres.

Tests indicate that the SRM algorithm remains highly efficient for low to intermediate particle volume fractions. At high volume fractions (≥ 0.82 for disks in 2D and approximately 0.6 for spheres in 3D), the random migration strategy becomes increasingly ineffective: particles tend to jam, and convergence can be achieved only by using very small swelling and migration rates. With an appropriate choice of control parameters, however, the SRM procedure is capable of generating dense configurations that exceed the classical Random Close Packing (RCP) limit in two dimensions (82% for monodisperse disks [27]). An example of an RCP-like configuration consisting of 5000 identical disks at a volume fraction of 87%, generated using the present method, is shown in Fig. 3a.

In three dimensions, the efficiency of the SRM algorithm is more limited. For monodisperse spheres, the practical upper volume fraction limit is approximately 60% (Fig. 3b). Beyond this threshold, the random migration mechanism becomes too inefficient to resolve overlaps, and more advanced algorithms, such as event-driven molecular dynamics, are required to generate configurations at higher densities.

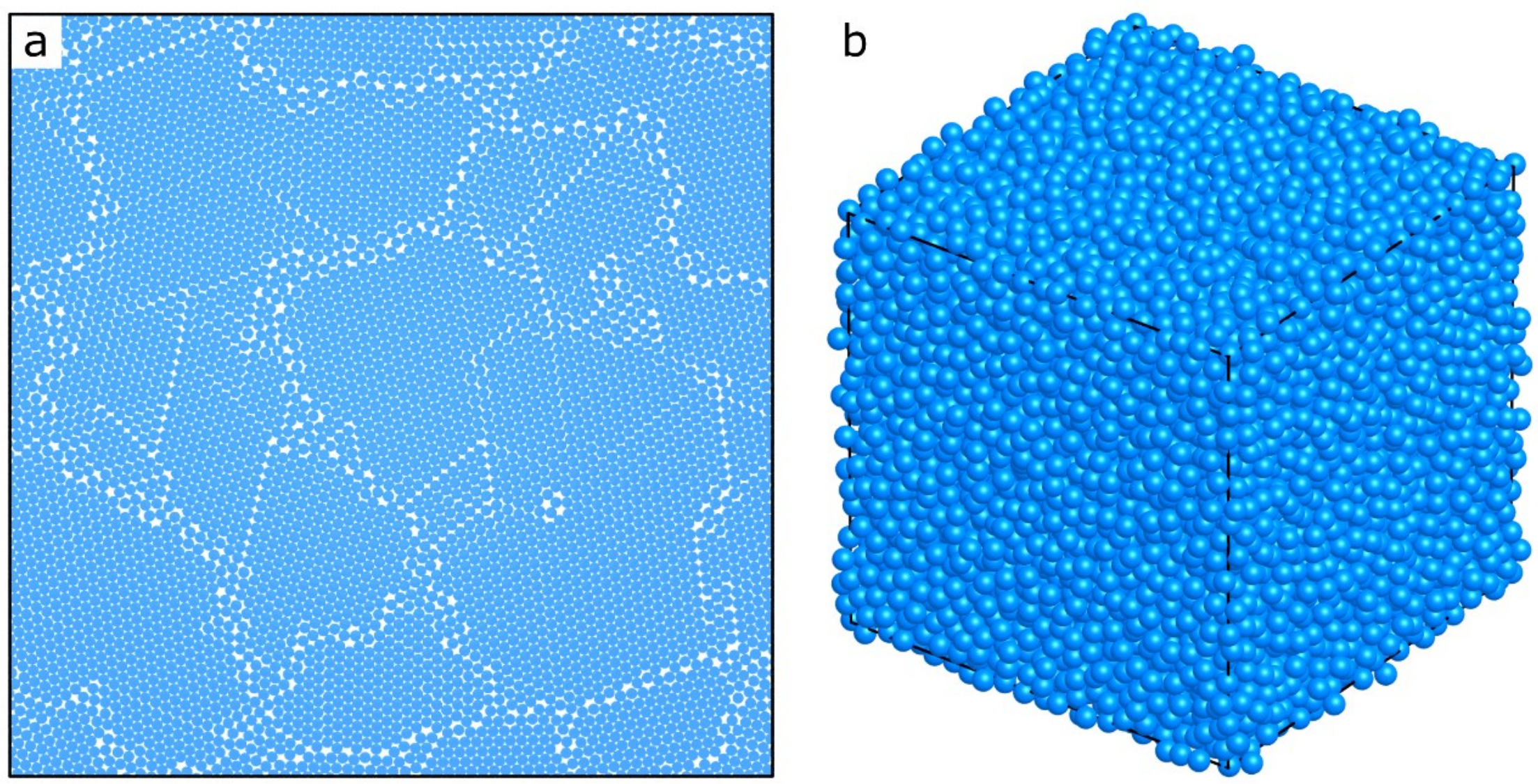

Fig. 3. Random close packing configuration of 5000 monodisperse disks with volume fraction of 87% (a) and 10000 spheres with volume fraction of 60% (b).

### 3.2. Generation of non-equilibrium distributions

The SRM algorithm was compared with the widely used event-driven molecular dynamics (MD) algorithm of Lubachevsky and Stillinger [2]. A key advantage of the SRM approach is its ability to generate a broad spectrum of particle distributions, from strongly clustered to equilibrium-like, within a single unified framework and using only a small number of control parameters. In contrast, the MD algorithm is inherently best suited for producing equilibrium or near-equilibrium configurations. Although non-equilibrium structures can be obtained in MD simulations by prescribing a large particle-growth rate (Fig. 4b), the resulting configurations remain significantly closer to equilibrium (Fig. 4a) than to the clustered microstructures generated by the SRM algorithm (Fig. 4c). This distinction is clearly reflected in the corresponding nearest-neighbor distance distributions shown in Fig. 4d, where the SRM-generated configuration exhibits the characteristic signatures of clustering that are absent in the MD-based results. Several RVEs were generated for each case, and average results are presented in Fig. 4d.

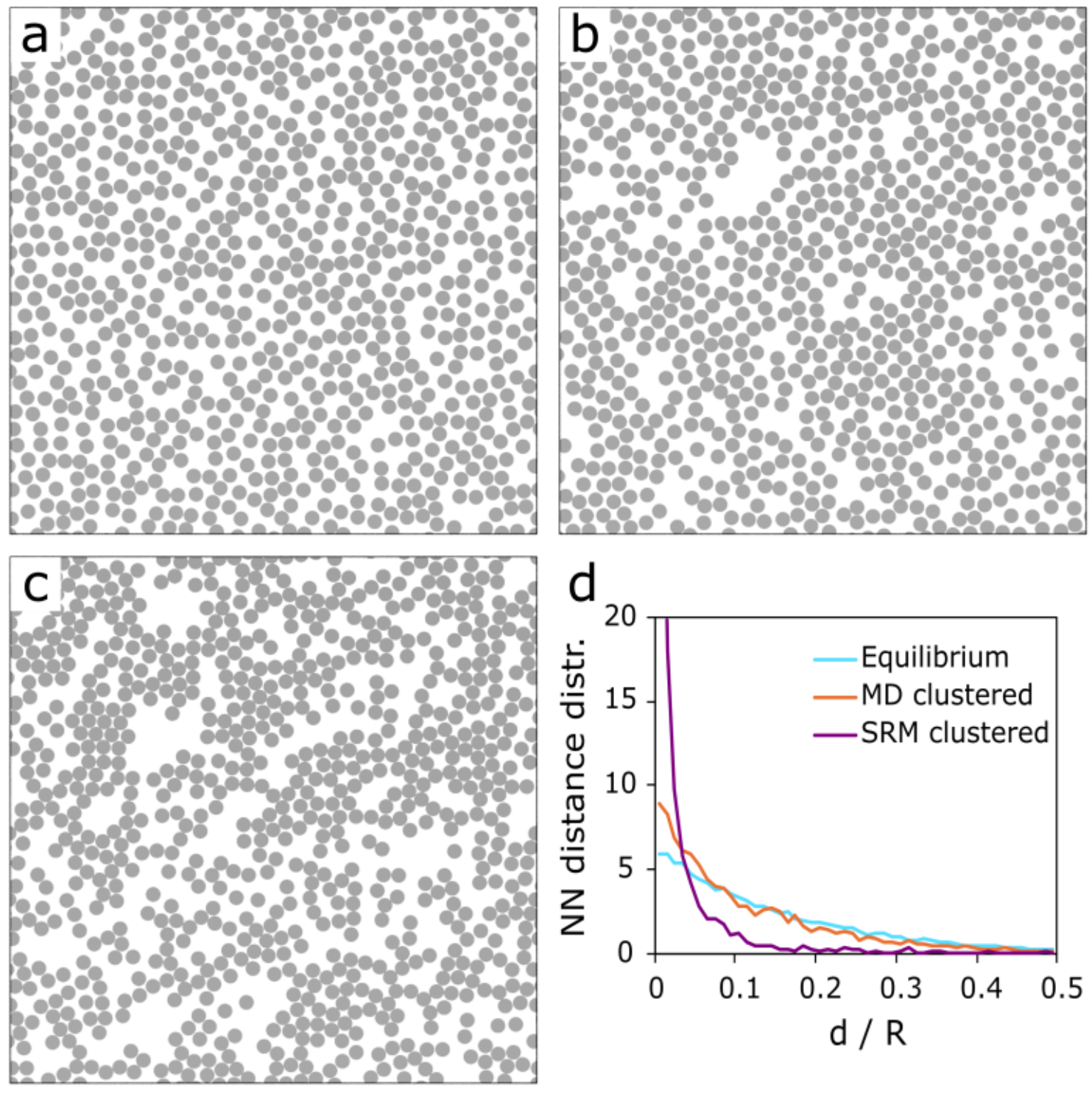

Fig. 4. Comparison of equilibrium distribution of fibers with volume fraction of 0.5 (a) with non-equilibrium distribution generated by MD algorithm (b) and clustered distribution generated by SRM algorithm (c), and corresponding nearest neighbor distance distributions (d).

Fig. 5 compares the local volume fraction (LVF) distributions for equilibrium, MD-clustered, and SRM-clustered particle configurations. The LVF for each particle was computed as the ratio of the particle area to the area of its corresponding Voronoi cell. Because the standard deviation of the LVF distribution is highly sensitive to clustering [28], the results in Fig. 5 highlight the different degrees of clusterization produced by the SRM and MD algorithms. The SRM method generates markedly broader LVF distributions, reflecting stronger local heterogeneity and more pronounced clustering compared with the MD-based configurations.

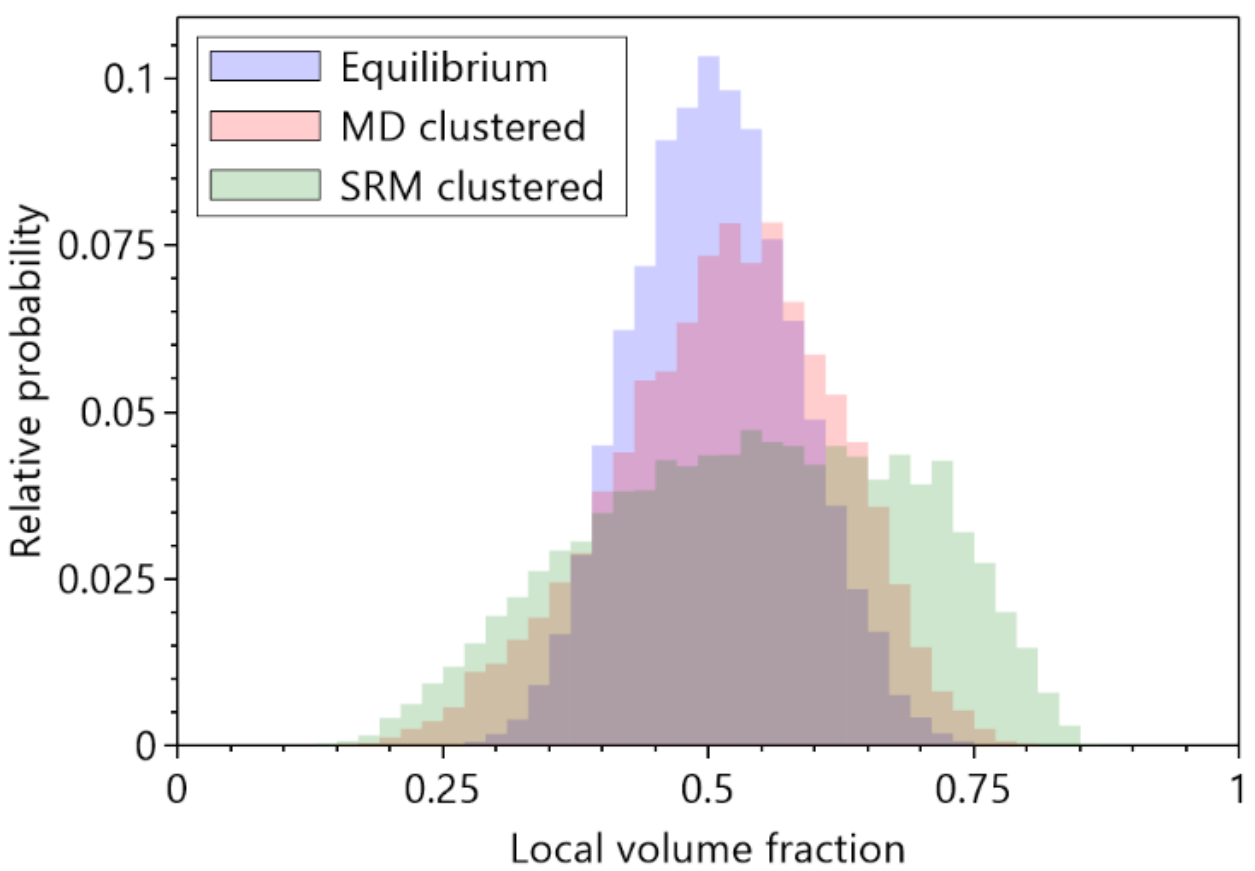

Fig. 5. Comparison of local volume fractions of equilibrium distribution with clustered distributions generated by MD and SRM algorithms for a global volume fraction of particles equal 50%.

In three dimensions, the visual distinction between clustered and equilibrium configurations is less pronounced than in two dimensions. Nevertheless, the clustered arrangement of 5000 identical spheres at a volume fraction of 50%, generated using the SRM algorithm, exhibits a noticeably more heterogeneous spatial distribution, with clearly identifiable matrix-rich regions (Fig. 6a). In contrast, the equilibrium configuration (Fig. 6b) appears more uniform, with fewer extended low-density zones. These differences become even more apparent for the LVF distributions shown in Fig. 7, where the SRM-generated clustered configuration displays a broader LVF distribution. LVF distributions obtained using the MD method for different particle growth rates were practically indistinguishable from those of the equilibrium configurations.

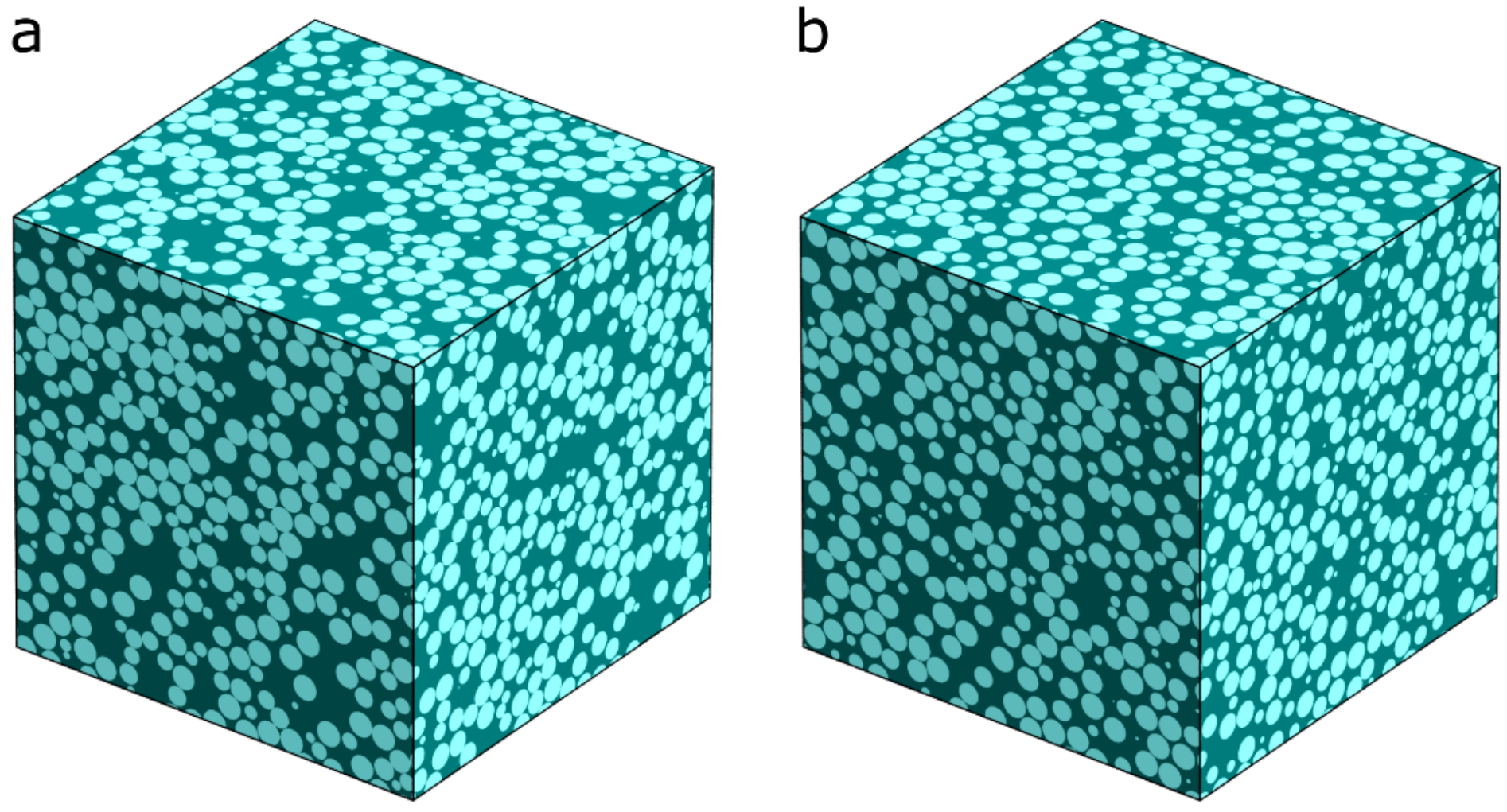

Fig. 6. Examples of RVE showing a clustered configuration (a) and an equilibrium configuration of 5000 identical spheres at a volume fraction of 50% (b).

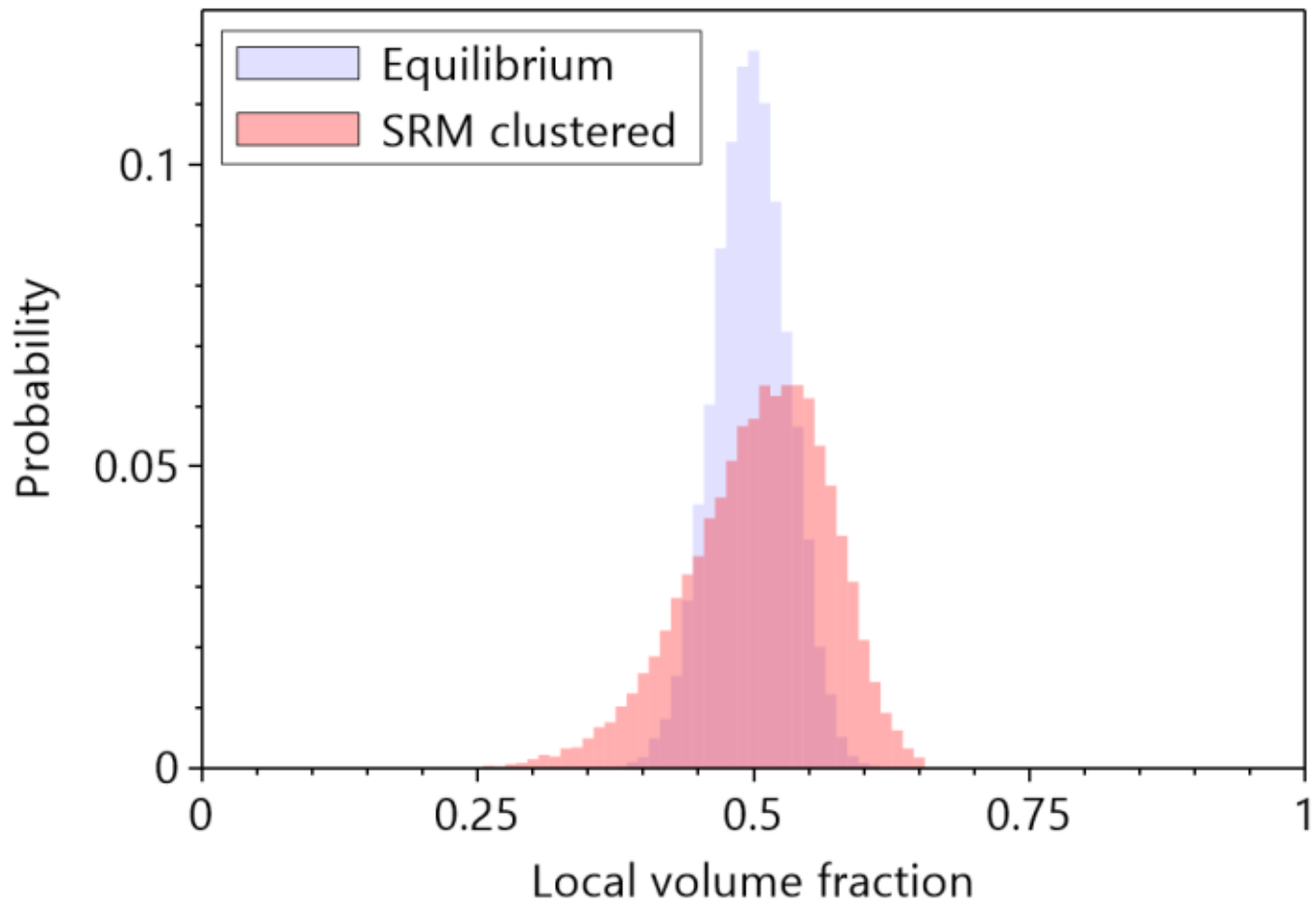

Fig. 7. Local volume fraction distributions for equilibrium and SRM clustered configurations of 5000 spheres at a volume fraction of 50%.

## 3.3. Application example

The SRM algorithm can be readily extended to a wide range of filler geometries [29], since the only geometry-dependent operation required is the collision detection routine, for which efficient algorithms exist for many particle shapes. As an illustrative example, the percolation behavior of suspensions of thin circular platelets, modeled here as spherodisks, was considered.

To investigate how platelet arrangement influences percolation behavior, a series of RVEs with varying inclusion volume fractions was generated. Each RVE contained 20 000 platelets with aspect ratios of 100 and 200, and number densities ranging from 0.6 to 5. The non-dimensional number density $\rho$ is defined as

$$\rho = \frac{ND^3}{L^3} \tag{1}$$

where $N$ is the total number of particles, $D$ is the platelet diameter, and $L$ is the characteristic size of the RVE. This non-dimensional formulation enables direct comparison of percolation behavior across systems with different particle sizes and domain dimensions.

The generation parameters were initially selected to produce house-of-cards (HoC) structures. By choosing a large swelling rate and a low migration rate, the SRM algorithm produces dense platelet networks in which many particles meet at diverse orientations (Fig. 8a). Such structures resemble the classical “house-of-cards” morphologies observed in colloidal suspensions, where platelet edges and faces carry different surface charges [30]. These interconnected configurations were then relaxed using zero swelling rate and a high migration rate to obtain metastable, partially ordered arrangements, yielding metastable microstructures containing large “quasi-nematic” domains (Fig. 8b). In Fig. 8, particle colors encode platelet orientation, with similarly oriented platelets assigned similar colors. The contrast between the two cases is evident: in Fig. 8a, neighboring platelets exhibit mostly random orientations with numerous edge-to-face contacts, whereas in Fig. 8b, extended regions of nearly parallel platelets emerge.

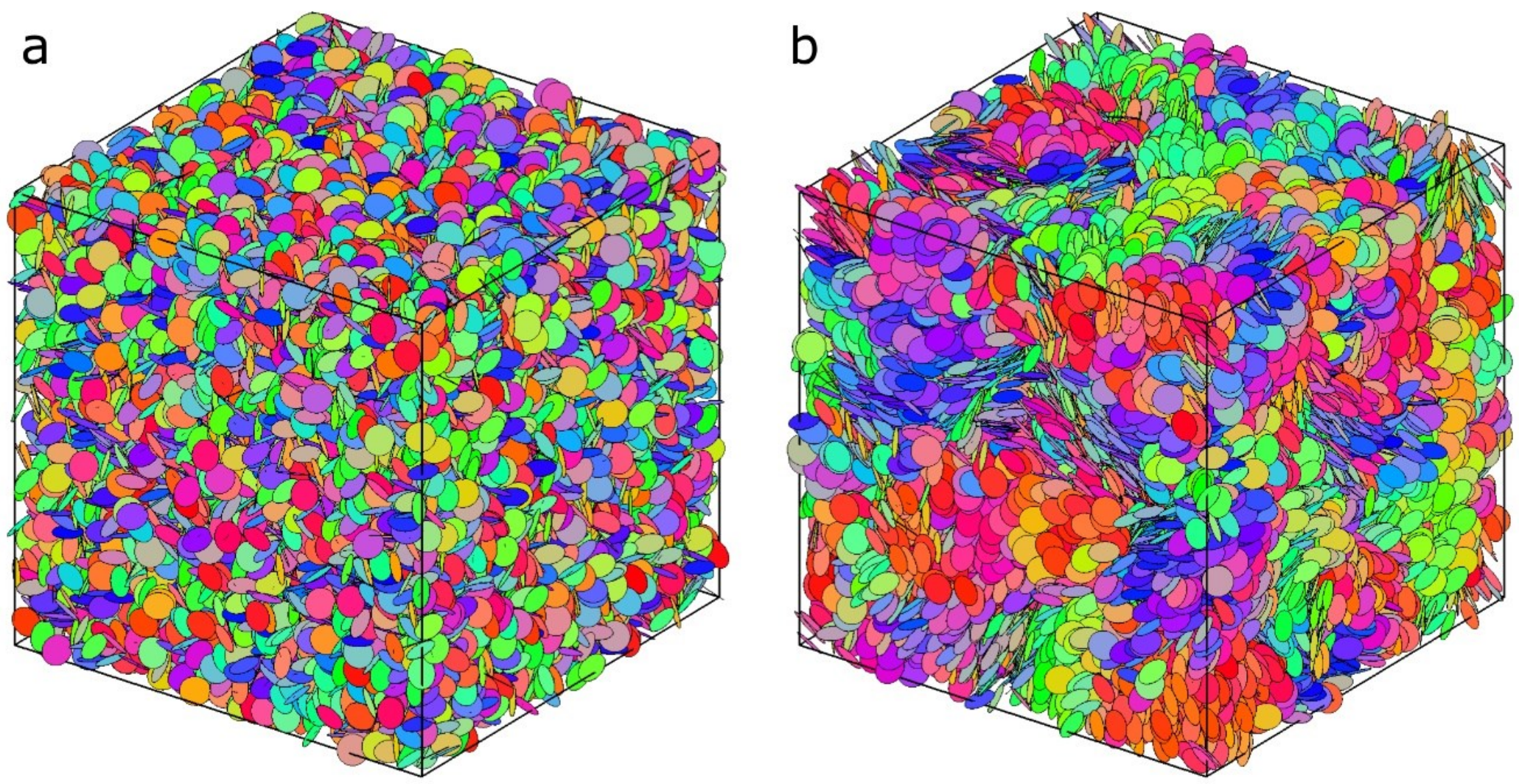

Fig. 8. Examples of generated “house-of-cards” (a) and “pseudo-nematic” (b) configurations of 20000 thin platelets with non-dimensional number density equal to 5 (periodic image particles on RVE boundaries are not shown for clarity).

For each generated microstructure, the critical percolation distance was computed using the hard core–soft shell model [31]. This distance is defined as the minimum tunneling gap required for the formation of a percolating cluster that spans the RVE in at least one spatial direction. The results are summarized in Fig. 9, which presents the normalized critical percolation distance $\delta_c/D$ as a function of platelet number density. The results indicate that HoC-type structures emerge at number densities above approximately 0.6, and the corresponding critical percolation distance decreases sharply for these highly interconnected platelet arrangements. In contrast, for the relaxed configurations, where quasi-nematic domains develop at higher number densities, the critical percolation distance is roughly an order of magnitude larger than that of the HoC structures. This pronounced difference highlights the strong influence of microstructural arrangement on percolation behavior and underscores the importance of accounting for particle organization in structure–property analyses. The critical percolating distances for the RSA-generated configurations are also included in Fig. 9 for comparison, and they show good agreement with the values reported in [32].

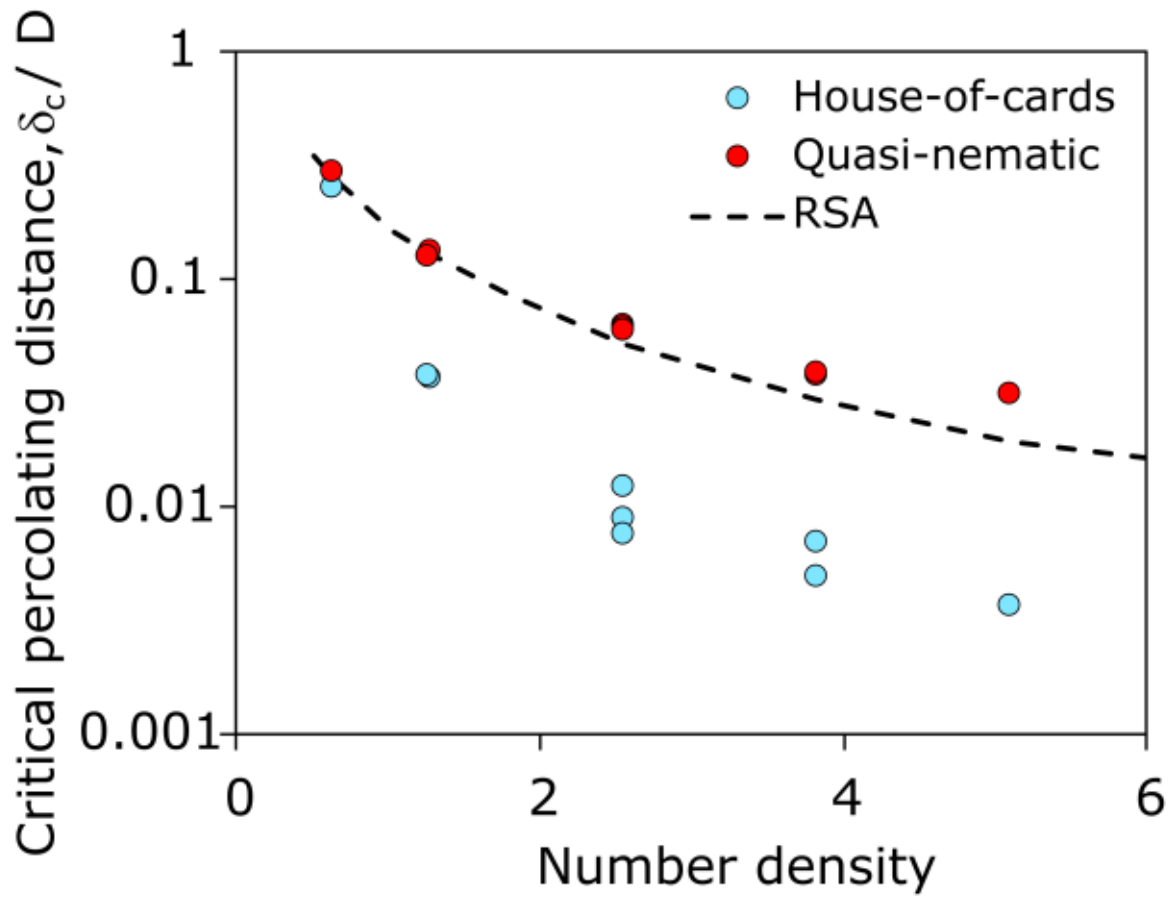

Fig. 9. Normalized critical percolating distance as a function of platelet number density.

During the further densification process, the initially loose quasi-nematic domains undergo progressive compaction. As the inter-domain spacing decreases, platelets are forced into closer proximity, and local rearrangements lead to the formation of stacked, face-to-face platelet assemblies. This transition reflects a shift from loosely aligned domains to densely packed, mechanically constrained structures characteristic of high-volume-fraction suspensions. Fig. 10 illustrates this process for a non-dimensional number density of $\rho = 30$, obtained from a configuration dominated by large quasi-nematic domains (Fig. 8b).

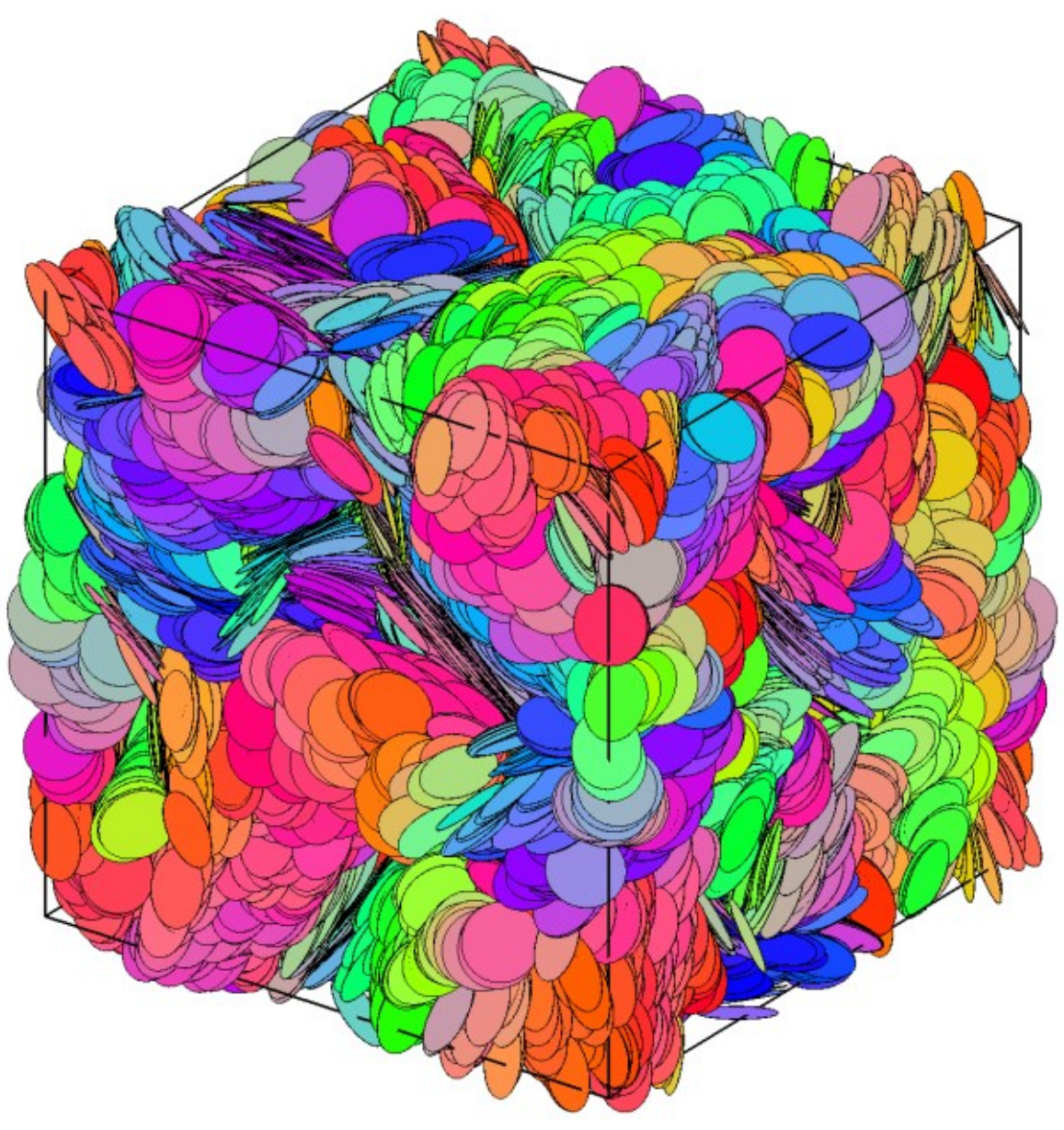

Fig. 10. Evolution of a platelet configuration with large quasi-nematic domains into a more compact structure with stacked clusters at a non-dimensional number density equal to 30.

This example demonstrates the versatility of the SRM algorithm in generating qualitatively different microstructures, from highly disordered, interconnected networks to partially ordered, domain-like arrangements. Such flexibility makes SRM a powerful tool for investigating structure–property relations in composite materials and colloidal suspensions exhibiting non-equilibrium particle distributions.

## 4. Conclusions

This work presented the Swelling and Random Migration (SRM) algorithm for generating statistically representative microstructures of particulate composites. The SRM framework provides a flexible and computationally efficient approach for producing a wide range of particle arrangements, from equilibrium-like to strongly clustered configurations, within a single unified methodology. Its ability to resolve multiple overlaps through collective rearrangements, combined with an efficient cell-based neighbor-search strategy, enables near-linear scaling for low to intermediate particle volume fractions, allowing simulations with up to $10^7$ particles in both two and three dimensions.

A key advantage of SRM over classical event-driven molecular dynamics (MD) algorithms lies in its capacity to generate non-equilibrium microstructures. While MD is inherently suited for equilibrium or near-equilibrium configurations, SRM can reliably produce highly interconnected, clustered, or metastable arrangements by appropriate tuning of swelling and migration rates. This versatility is essential for studying composite materials and colloidal suspensions, where non-equilibrium structures frequently arise.

The algorithm's extensibility was demonstrated through its application to thin circular platelets. By adjusting control parameters, SRM successfully generated both "house-of-cards" structures and quasi-nematic domains, enabling systematic investigation of how platelet organization influences percolation behavior. The resulting analysis showed that microstructural arrangement has a profound effect on the critical percolation distance, with HoC configurations exhibiting significantly lower thresholds than relaxed, domain-like structures.

The SRM algorithm provides a robust and flexible tool for generating complex particle distributions in periodic RVEs. Its ability to mimic diverse microstructures observed in real materials makes it particularly useful for structure–property studies, percolation analysis, and the development of advanced composite and colloidal systems. Although the SRM algorithm provides an efficient framework for generating a wide range of particle microstructures, it is important to emphasize that the method is purely geometrical. The particle rearrangements are not governed by physical forces. As a result, SRM does not reproduce the true physical evolution of particulate systems and cannot be used to study phase transitions or dynamic processes that depend on interparticle potentials.

**Acknowledgement**

This research was funded by the Latvian Council of Science, grant No. ES RTD/2022/16 "Graded Interphases for Enhanced Dielectric and Mechanical Strength of Fiber Reinforced Composites" carried out under the M ERA.NET 3 scheme (European Union's Horizon 2020 research and innovation program under grant agreement No 958174).

**Declaration of competing interest**

The authors declare that they have no known competing financial interests or personal relationships that could have appeared to influence the work reported in this paper.

**Data availability**

Reference implementation of the algorithm is available on GitHub (https://github.com/RiSearcher/SRM).